\documentclass[
  journal= ,
  manuscript=article-type,
  year=,
  volume=,
]{cup-journal}

\usepackage{amsmath}
\usepackage[nopatch]{microtype}
\usepackage{booktabs}
\usepackage{caption}

\title{`Congratulations, morons': Dynamics of Toxicity and Interaction Polarization in the Covid Vaccination and Ukraine War Twitter Debates}

\author{D.S. Axelrod}
\affiliation{Informatics, Indiana University, Bloomington, 47405, Indiana, USA}

\email[F. Author]{daaxelro@iu.edu}

\author{B.P. Harper}
\affiliation{Informatics, Indiana University, Bloomington, 47405, Indiana, USA}

\author{J.C. Paolillo}
\affiliation{Informatics, Indiana University, Bloomington, 47405, Indiana, USA}


\begin{document}

\begin{abstract}
The existence of polarization and echo chambers has been noted in social media discussions of public concern such as the Covid-19 pandemic, foreign election interference, and regional conflicts. However, measuring polarization and assessing the manner in which polarization contributes to partisan behavior is not always possible to evaluate with static network or affect measurements. To address this, we conduct an analysis of two large Twitter datasets collected around Covid-19 vaccination and the Ukraine war to investigate polarization in terms of the evolution in influencer preferences and toxicity of post contents. By reducing retweet behavior in each sample to several key dimensions, we identify clusters that reflect ideological preferences, along with geographic or linguistic separation for some cases. By tracking the central retweet tendency of these clusters over time, we observe differences in the relative position of ideologically unaligned clusters compared to aligned ones, which we interpret as reflecting polarization dynamics in the information diffusion space. We then measure the toxicity of posts and test if toxicity in one cluster can be temporally dependent on its structural closeness to (or toxicity of) another. We find evidence of ideological opposition among clusters of users in both samples, and a temporal association between toxicity and structural divergence for at least two ideologically opposed clusters in our samples. These observations support the importance of analyzing polarization as a multifaceted dynamic phenomenon where polarization dynamics may also manifest in unexpected ways such as within a single ideological camp. 
\end{abstract}

\section{Introduction}

The last few years have witnessed major geopolitical upheavals, notably a global epidemic that caused millions of deaths, and two major regional wars in Eastern Europe and the Middle East launched less than two years apart. Consuming news on these events occurs in the context of existing ideological frameworks, political tribalism, and media fragmentation. This has in many cases produced difficulties for political systems where citizens find it increasingly difficult to identify a shared basis for knowledge and understanding of current events. The challenge this poses, especially for democratic systems, and the ability to observe large-scale discussions on social media has prompted renewed interest in polarization on social media (\cite{McCoy, bail}).    

Previous work on polarization has provided insights including a distinction between political (\cite{DiMaggio, Fiorina}) and affective polarization (\cite{Iyengar, Kubin, Zwinkels}). Political polarization refers to situations where individuals differ and sort themselves according to political beliefs. In contrast, affective polarization reflects the extent to which individuals evaluate opposing political groups in a negative emotional frame. However, as related social phenomena, the distinction between these two models of polarization may not always be clear, and previous research has found evidence that the two processes are related in some close way (\cite{Zwinkels}). There is a simple plausibility that the emotional evaluation of other political groups will be influenced by one's perception of their political views. Similarly, one's emotional or moral framing of others may inhibit or facilitate the ease with which one accepts views from others. Yarchi et al. have also identified what they refer to as interactional polarization on social media platforms, referring to a lack of interaction between opposing user groups on social media (\cite{Yarchi}). Although helpful as a distinct manifestation of polarization in the context of social networks, interactional polarization may itself be the result of interplay between political or affective polarization and homophily dynamics within networks. 

In the literature, empirical analyses of polarization often treat it as a static phenomenon: one inspects a network for specific, quantifiable properties that one believes to reflect polarization. Alternatively, just a few snapshots of changing dynamics are investigated (\cite{con_pol, jiang2021social, PolPolar2023,ideol_infer}). These approaches encounter methodological problems that arise from incomparability of network data collected via different procedures and comparing networks of disparate sizes. One approach is to identify and measure a choke point (i.e. a structural hole) in a network, or use measures that reflect how easily one can navigate across a network (\cite{gen_eucl}). However, such approaches are not ideal for measuring polarization in systems with more than two poles, and the presence of choke points or other characteristics of network structure such as modularity can be highly biased by commonly-encountered dataset properties. Perhaps the most common example of this issue is the presence of spam or bot posts (\cite{Barbera_2}). Using these posts when constructing and analyzing a network can bias one's sense of polarization, especially when these users act in atypical ways, e.g., by retweeting users with conflicting positions. For this reason, a common approach is to restrict the network inclusion criteria to users already known to be politically active and relevant. However, if the inclusion criteria are not well informed, an analysis may miss important groups and cleavages among the studied population.

Even when filtering a network to retain context-relevant relationships, a static measure of polarization neglects an important component of polarization as experienced by people communicating on a social media network, namely change in or reification of ideological positions over time. In other words, one needs to look at the temporal dynamics of a social network to fully understand its polarization. Within Twitter, various kinds of network linkage are present, with retweeting behavior accounting for the most numerous kind of linkage. As network positions evolve on Twitter, the structure of retweeting behavior changes: different influencers increase or decrease in their popularity, and different sets of users retweet them. Hence, polarization in a Twitter network is partly expressed via changes in its retweeting structure. These changes can be observed as movement in the network proximity of user clusters in a latent space. In addition to their numeric prominence, retweets also have the characteristic of being relatively transparent for interpretation compared to replies or quote tweets, as both latter interaction forms are used in more heterogeneous ways to engage with both aligned and adversarial users.

In cases where a filtered network exhibits high fragmentation over time, divisions among users may still not in themselves reflect the presence of animosity between groups, which we argue is a critical aspect of polarization. A tendency toward homophily-based relationships may, for example, link users sharing similar niche interests. This may be reinforced further when there is a geographic dependence, as in the example of sport league federations. Since network structure may not in itself be a sufficient indicator of polarization, we add an additional condition in our search for evidence of polarization. Informed by previous work on affective polarization (\cite{affect_pol}), we hypothesize that the affective states of users in a polarized environment can play a role in political polarization dynamics. For example, provocative content shared by one cluster, marked by insults or ad hominem attacks, may result in provocative responses by other clusters. Alternatively, clusters may diverge in network terms due to inciting incidents that impact influencer preferences. Hence, it is important to observe the direction and timing relationship of network changes alongside affective qualities of the discourse.

\section{Data \& Methods}
 
\begin{table}
    \centering
    \begin{tabular}{p{0.1\linewidth} | p{0.4\linewidth} | p{0.4\linewidth}}
    \toprule
    Sample & Original Query & Secondary Query  \\
    \midrule
    Covid & vaccine, pfizerbiontech, covid19vaccine, oxfordvaccine, pharmagreed, sputnikv, covidiots, getvaccinated, thisisourshot, vaccinessavelives, greatreset, mybodymychoice, iwillnotcomply, endthelockdown, kungflu, plandemic & nazism, holocaust, holodomor, genocide, fascism, hitler, stalin, communism, soviet, world war, cold war ww2, imperialism, authoritarianism, marxism, capitalism\\
    \midrule
    Ukraine & ukraine, russia, putin, moscow, kremlin, nato, luhansk, donetsk, kyiv, kiev, zelensky, fsb, kgb, slava ukraini, donbas& Same as above\\
    \bottomrule
    \end{tabular}
    \caption{Terms used in the original data collections and the common secondary query used for user inclusion criteria to index ideologically engaged users. For space considerations we provide one example per noun, omitting terms containing the same root or compound terms used in the Covid vaccine data collection.}
    \label{terms_table}
\end{table}

This analysis employs two Twitter datasets. One is a published dataset centered on Covid-19 vaccination (\cite{deverna2021covaxxy}) spanning the period from January 4, 2021 to February 7, 2023; the other is a dataset on the Ukraine war collected similarly, spanning January 28, 2022 to December 2, 2022 (\cite{whitepaper}). Both episodes reflect prominent US and global political divisions, whose positions around these issues seemed to grow more entrenched. Speaking in terms of the American left-right political spectrum, the political right consolidated around opposition to vaccination and other public health mandates favored by the left. In the Ukraine conflict, members of the right favored Russia while the center of gravity on the left favored Ukraine. Because left-right divisions in the US do not map neatly to those in Europe and elsewhere, the story becomes more complex in the global context. Nonetheless, Covid vaccination and the Ukraine war have prompted people to sort themselves into opposing camps, which suggests we may find evidence of polarization expressed in these Twitter samples. 

\begin{figure}[h!]
    \centering
    \includegraphics[width=1\textwidth]{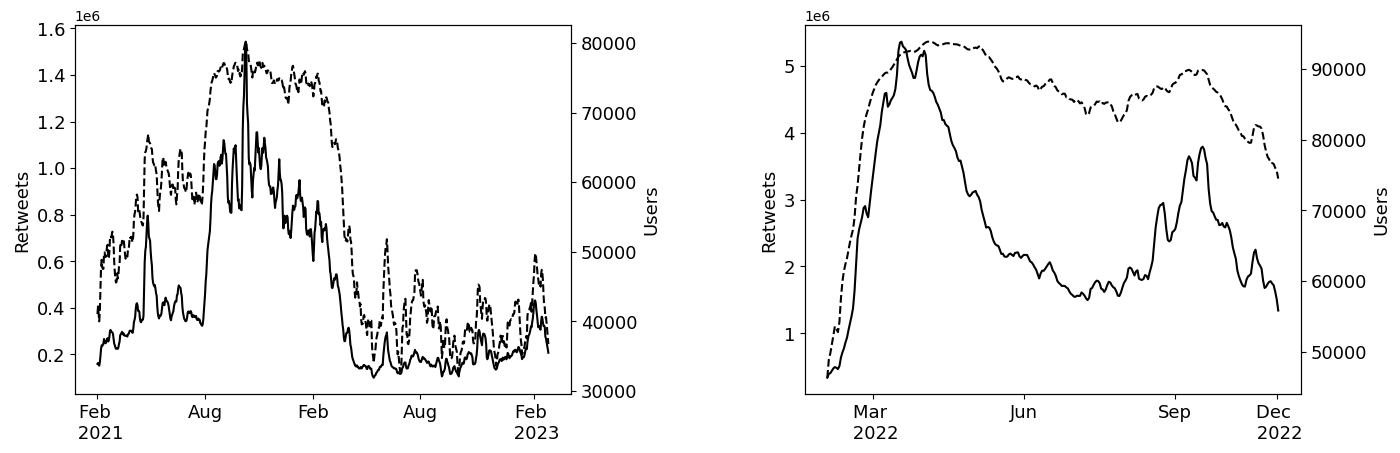}
    \caption{Time series for Covid (left) and Ukraine (right) samples. Retweet counts are depicted with solid lines while the number of unique users is shown with the dashed curves. To account for differences in scale, each plot has separate y-axes for each variable.}
    \label{fig:times}
\end{figure}

Both samples began with queries to the Twitter API starting with limited set of topic-relevant terms e.g., "Ukraine" and expand the query iteratively by examining novel frequent term in the collected data before fixing the query for the duration of data collection. The resulting datasets are dominated by news content and spam, as well as topics adjacent to the main query theme. This presents a challenge for studying polarization. Although adjacent topics and news sharing can reflect polarization dynamics among users, a dataset dominated by these behaviors will reflect a picture of polarization that captures non-ideological divisions, a problem that even stringent filtering may not be able to fully address. For this reason, previous approaches to polarization in social networks restrict their analysis to known political influencers or use some other heuristic to exclude irrelevant network relationships (\cite{barbera2015birds}). To address this concern but without limiting ourselves to a predetermined set of influencers, we leverage insights from previous analysis of ideology in the data (\cite{contrary}) to identify a set of concepts that index interaction with prominent ideological frameworks. Many of these concepts are informed by the global experience of the Second World War and Cold War and relate to cultural expressions of this legacy such as Godwin's Law. To ensure persistent engagement with ideological content, our inclusion criteria require users to retweet or post at least 28 (Ukraine) or 7 (Covid) posts with one of the ideological terms during the Ukraine (305 days) and Covid (761 days) sample periods. This inclusion criterion results in a comparable number of users for the Covid (101,036 retweeters, 9,015 influencers) and Ukraine (101,990 retweeters, 9,661 influencers) samples. We then consider all posts created and retweeted by our user sets as captured in the original data collection queries, having greater confidence that other content shared by these users is relevant to ideological discourse compared to the nature of the full datasets. This results in 33,287,386 retweets in the Covid-19 data and 75,068,498 retweets in the Ukraine data.

\section{Analysis}

To study changes in polarization over time, we consider two features in retweet behavior that may reflect changes in political or affective polarization. For political polarization, we examine retweeting preferences in influencers, with the understanding that these can reflect political identities (\cite{conover2011predicting}). For our samples, the ideological character of each cluster is then interpreted through its prominent hashtags and posts. We furthermore consider the structural position of clustered users via a rank-reduced approximation of the retweet matrix, which has also been constructed to preserve its time component, to allow for the network's evolution to be observed as per-cluster time series. These are compared to similar per-cluster time series for toxicity using Granger causality. These analysis steps are explained in turn below. 

\subsection{Network structure of retweeting}

Retweets of influencers by users are the basic form of link we employ. The time dimension is addressed by segmenting the sample by the timestamp of each retweet into overlapping 7-day windows; windows are defined for each calendar day in a sample. An incidence matrix is constructed for each window recording retweets of each influencer by the union of all retweeting users in that window, represented as rows. Window incidence matrices are double-centered (by row and column) to address the relative propensity of influencers and users in retweeting, before subjecting them to Principal Component Analysis (PCA) by way of Singular Value Decomposition (SVD) as in equation \eqref{pc_space}; for each window, we retain the first 30 dimensions as scores, computed as $A\widetilde{Q}$, where $\widetilde{Q}$ is the reduced-rank rotation matrix. The score matrices for each window are concatenated into the sample matrix $S$ for further SVD after column-wise $z$-score centering and scaling. 

\begin{equation}
    A = P \Lambda Q^T
\label{pc_space}
\end{equation}
This arrangement preserves the scaling and linearity of the relationship between sample principal components (henceforth, simply components) and those of windows, permitting straightforward projection of the scores for users within specific windows. Retention of  sample components from the sample PCA is guided by scree and PC plot inspection. This approach also does not fix the position of influencers over time, allowing their roles to change as their views and the discussion evolves.

Clustering users in the reduced information diffusion space is accomplished using HDBScan (\cite{hdbscan}), a density based approach to clustering that also assigns observations to an outlier group. This allows us to focus on clusters of users that clearly exhibit common retweet behaviors. Prior to clustering, user vectors are $L2$-normalized so as not to differentiate users by their activity levels. Initially, we fit our clustering model to users exceeding a Euclidean distance threshold from the origin (10). We then use this model to predict the cluster membership for the remainder of users. This procedure ensures that clustering is not overly influenced by inactive users, avoiding clusters centered  near the origin that do not engage with the main retweet trends and instead assigning these inactive users to the outlier group. 

Clustering and visualization of the user clusters in the sample space identifies relevant groupings that may differ in ideological terms, for which the hashtags and usernames are important cues, as well as specific retweets sampled from within each cluster. This type of observation corresponds to political polarization observed via static means. 

\subsection{Time series}

Cluster positions over time are computed as per-window average centroids over all users within a cluster. To compare clusters, we compute their structural dissimilarity pairwise between clusters as in equation \eqref{struct_sim}.  
\begin{equation}
 D_{c1,c2} = -cos(\mu_{c_1},\mu_{c_2})
\label{struct_sim}
\end{equation}
The paths of cluster centroids through the PC space provide an indication of how their network positions change over time, i.e., via churns and exchanges of influencers favored by different groups of users at different times. If clusters diverge in the PC space, they share fewer influencers or retweet more disjoint sets of them. Such changes potentially correspond to dynamic observations of political polarization. 

To address affect-related mechanisms in polarization dynamics, we consider toxicity as a proxy variable, measured using Google's Perspective API (\cite{Perspective}). Under the assumption that periods of affective polarization should exhibit language broadly considered toxic by this model, we computed toxicity for all posts retweeted at least 10 times in our samples. We furthermore inspected posts with highly toxic scores to confirm the appropriateness as a proxy variable. A cluster's toxicity was aggregated from constituent post scores according to equation \eqref{ind_toxicity} ; this preserves the interpretation of toxicity given in the API documentation, according to which the toxicity score is the likelihood that a user would find something toxic in a post. The aggregate version can be thought of as the probability that toxicity will be encountered within a cluster by someone randomly engaging with its constituent users. For pairwise comparisons of toxicity between two clusters $c1$ and $c2$, we also compute the likelihood of encountering toxicity in either cluster according to equation \eqref{joint_toxicity}. These values are computed window by window, producing time series.

\begin{equation}
 T_c = 1- \prod_{i \in c}{[1-F_{i} \cdot T_{i}]}
\label{ind_toxicity}
\end{equation}

\begin{equation}
 T_{c1,c2} = 1-(1-T_{c1}) \cdot (1-T_{c2})
\label{joint_toxicity}
\end{equation}

\subsection{Temporal association}

When considered individually, the toxicity and structure time series could function as independent models of polarization dynamics for our samples. In the case of toxicity, one may first investigate if there is a dependency between the toxicity of one cluster and another. In the context of political discussions, finding a pattern where toxicity in one cluster precedes toxicity in another could indicate of a negative affective relationship between the clusters. A similar line of inquiry can be extended to both toxicity and structure variables to test if the structural relationship between clusters depends on the affective states of those clusters, or vice versa. A common approach from econometrics to test for such temporal relationships is known as the Granger causality test (\cite{Granger}). Granger bivariate tests evaluate if a linear autoregressive model for one variable is improved by the inclusion of lagged values from another variable.

Because Granger tests require time series to be stationary, we remove trends and periodicity from our series by fitting an ordinary least squares model over rolling windows containing the preceding month's observations. In each iteration, we only retain the residual for the final observation, ensuring that inferred trends are not biased by future observations. By restricting our de-trending procedure to historical data, we remain consistent with Granger test assumptions.

\section{Results}
\subsection{Sample spaces and clusters}
Scree plots and inspection of the PC plots yielded four PCs for both samples; subsequent cluster analysis resulted in four clusters for the Covid sample (C1, 32,766 users; C2, 5,875; C3, 11,601; and C4, 32,116) and five for the Ukraine sample (U1, 5,844 users; U2, 20,242; U3, 7,889; U4, 22,918; and U5, 17,663). The PC scores are displayed in pairwise PC plots in Figures \ref{fig:cov_clust} and \ref{fig:ukr_clust}; users are colored by clusters in each, corresponding to the coloring in the bar-plot key in the upper right of each figure. The diagonal plots give the univariate distributions of each cluster on each PC. 

\subsection{Toxicity}
Prior to investigating toxicity dynamics, we first test for differences in the clusters' toxicity probabilities. If one cluster is found to be substantially more toxic, this could help us understand the role that toxicity plays in polarization dynamics. We evaluate this with one-sided Mann-Whitney test (\cite{mannWhitney}) on the toxicity score distributions of all retweets per cluster. A significant p-value indicates that the toxicity distribution of the first cluster's retweets is stochastically greater than the toxicity scores in the second cluster. We then use the U-statistic from each test to compute the Area Under the Receiver Operating Characteristic curve (AUC), corresponding to the likelihood that a randomly chosen value in the first distribution is equal to or greater than a random value in the second distribution. In the Covid sample, C1 is the most toxic, followed by C4, C3, and C2. Testing clusters in that order, we obtain AUC values of 0.53, 0.51, 0.53 and p-values less than or equal to 1e-50. While significant, the distributions do not differ radically. Still, it is notable that American-centric clusters C1 and C4 are more toxic than the British and Canadian clusters. In the Ukraine sample, U2 is the most toxic, followed by U1, U5, U3, and U4. Again presenting results in that cluster order, we obtain AUC values of 0.52, 0.50, 0.51, and 0.51, again with p-values less than or equal to 1e-50. Whereas in the Covid sample, the most toxic clusters were ideologically diverse, U2 and U1 are both firmly in the pro-Russia camp.

\subsection{Cluster comparison time series}

\begin{figure}
    \centering
    \includegraphics[width=1\textwidth]{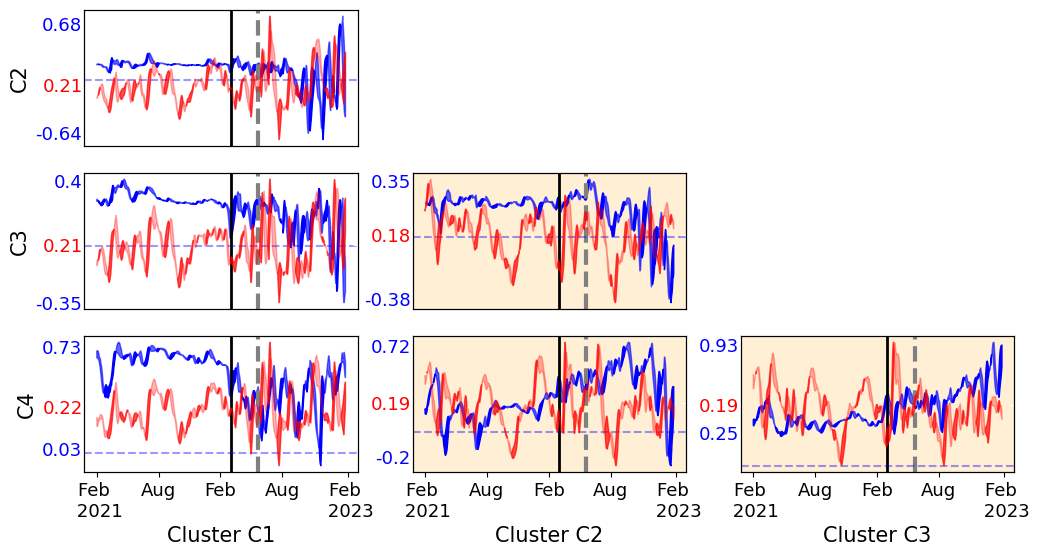}
    \caption{Time series plots for each pair of clusters in the Covid sample. Blue curves show the structural dissimilarity for those clusters as defined in equation \eqref{struct_sim} while red curves show the combined toxicity for the two clusters as defined in equation \eqref{joint_toxicity}. The shaded area hugging each curve corresponds to the difference between residuals from OLS fits and observations smoothed using a Gaussian kernel ($\sigma$ = 3). Each curve is plotted on its own y-axis with tick label colors corresponding to the respective curve colors. We include a light-blue horizontal dashed line to indicate the location of zero on the blue y-axis. A vertical black (solid) line indexes the beginning of the Ukraine war while a gray (dashed) vertical line shows where the time series is clipped for Granger tests. }
    \label{fig:cov_ts}
\end{figure}

Time series for toxicity and structural dissimilarity in the Covid sample are depicted in Figure \ref{fig:cov_ts} where cells are shaded when both clusters contain users opposed to public health mandates and vaccination. Additional details on each sample space and cluster can be found in the Interpretation subsection below. Series for structural dissimilarity (blue) show distinct motifs depending on whether the examined cluster pair is ideologically aligned. For example, between February and August 2021 aligned cluster pairs initially move apart from each other (up), then toward each other (down), before again converging. The inverse relationship is observed for this period for cluster pairs that are in opposition. Zooming out to a longer time scale, we observe that unaligned cluster pairs (first column) show an overall decrease in their structural dissimilarity in the first half of the sample, prior to entering a less stable period that coincides with the start of the Ukraine war and the public's shift in focus to that topic. The user exodus from the Covid mandate discussion is most clearly seen in the retweet and user counts shown in Figure \ref{fig:times}. In contrast, aligned cluster pairs C2,C4 and C3,C4 both show a fairly steady increasing trend in their structural dissimilarity past the sample midpoint and into October 2023. At this point C2,C4 diverges from this trend as C2 becomes less active. The magnitude of structural dissimilarity is higher for cluster comparisons involving cluster C4, suggesting especially idiosyncratic behaviors by that cluster.

Time series for toxicity also exhibit notable patterns. Common bursts in higher toxicity are found across series, suggesting that some aspect of series trends may be determined by circumstances and phases in vaccine distribution or mandate implementation. As with structural dissimilarity series, the second half of the sample exhibits less stability compared to the first. Motifs in these series again distinguish aligned cluster pairs from unaligned pairs. Once more focusing on the period between February and August 2021, we see that unaligned pairs exhibit an overall upward trend, while aligned clusters have a decreasing trend. During that same period, we also see examples where bursts in toxicity and structural dissimilarity line up.

\begin{figure}
    \centering
    \includegraphics[width=1\textwidth]{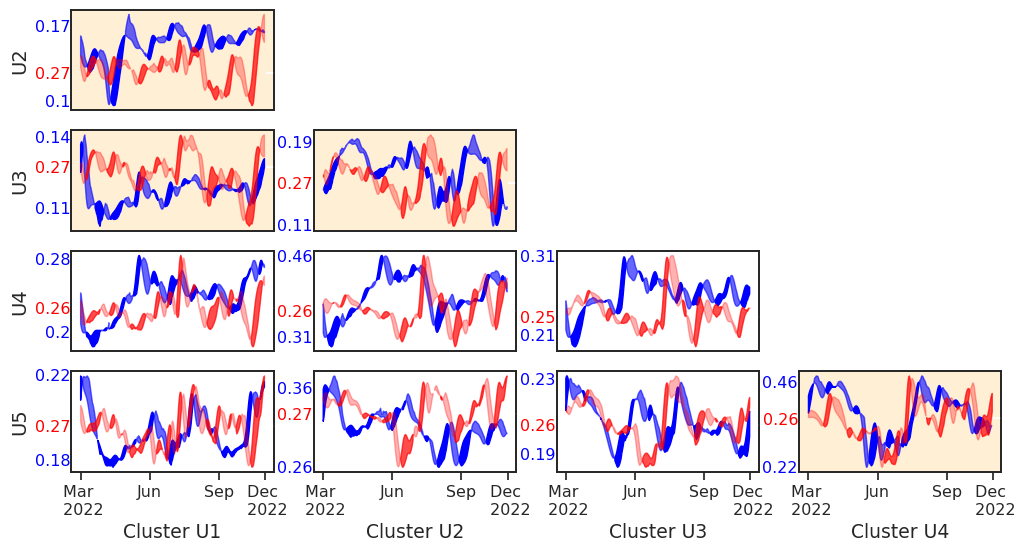}
    \caption{Time series plots for each pair of clusters in the Ukraine sample. Blue curves show the structural dissimilarity for those clusters as defined in equation \eqref{struct_sim} while red curves show the combined toxicity for the two clusters as defined in equation \eqref{joint_toxicity}. The shaded area hugging each curve corresponds to the difference between residuals from OLS fits and observations smoothed using a Gaussian kernel ($\sigma$ = 3). Each curve is plotted on its own y-axis with tick label colors corresponding to the respective curve colors.}
    \label{fig:ukr_ts}
\end{figure}

Time series for toxicity and structural dissimilarity in the Ukraine sample are depicted in Figure \ref{fig:ukr_ts} where cells are shaded when both clusters align with respect to their support for Russia or Ukraine. Motifs in the structural dissimilarity series are less clearly able to differentiate aligned from unaligned cluster pairs when compared to the Covid sample. Still, we do see such a result when comparing the series for U1,U4 and U4,U5. The former pair, being a pair of unaligned clusters, has an initial decreasing trend in structural dissimilarity during the first month of the war, followed by an increasing trend toward June, and a decreasing one toward August. The reverse of this pattern is found for aligned clusters U4,U5. As with the toxicity time series in the Covid sample, there are moments when bursts in toxicity are common to all pairs, notably around August (2022). With respect to the relationship between toxicity and structural dissimilarity series, the Ukraine sample does feature some moments when both peak at the same time but not as clearly or consistently as in the Covid sample. Interestingly, the toxicity and structure series for U4,U5 have similar and nearly aligned trend contours, in most contrast to the trends for U1,U4. The presence of contrasting temporal motifs for aligned and unaligned cluster pairs, especially in the Covid sample, suggest that we have managed to measure meaningful information about user groups with respect to both toxicity and and retweet structure.

\subsection{Granger Causality}

Having performed an initial inspection of the toxicity and structural dissimilarity time series, we test to see if there is statistical evidence for temporal dependencies between our clusters. Approaching polarization as a process suggests that predictive relationships may be especially insightful. Because each cluster can be characterized by its toxicity prevalence over time, one may ask if one cluster's toxicity predicts another cluster's toxicity. When introducing structural dissimilarity into this line of inquire, one may ask if a cluster pair's toxicity can predict the structural (dis)similarity of clusters in our sample spaces. To assess this, we test for Granger-causality between these variables. A significant result for these comparisons indicates that lagged values of one variable are informative in predicting future values of another beyond that second variable's own lagged values, which can be very cautiously interpreted as suggesting a possible causal relationship. 

As specified above, Granger tests are performed with respect to residuals of each time series in order to remove the trend components of a series. In addition, because the Covid sample enters a phase of low activity for the final 231 observations in the sample, we take a conservative approach and limit analysis to the first 500 observations. This helps us avoid spurious results coming from the noisy end period of the Covid sample. We test for temporal dependence between our variables in both directions, up to a lag of 155 observations in the Covid sample and 85 observations in the Ukraine sample.

When adjusting significance thresholds using a Bonferroni correction, three cluster pairs are found to exhibit Granger causality across our two samples. In the Covid sample, both an unaligned pair (C1,C4) and an aligned one (C3,C4) have significant results. In both of these cases, this is true for toxicity predicting structural changes and vice versa. While this result for C1,C4 maps well with our previous expectations, significant results for two anti-mandate clusters C3,C4 are unexpected. However, it is helpful to note that C3,C4 exhibit the strongest structural dissimilarity among all cluster pairs in the Covid sample. This raises the possibility that other forms of polarization are present in this sample. For the Ukraine sample, U1,U4 return significant results for all tests. This suggests a temporal dependency between the propensity for toxicity in each cluster, as well as a relationship between their joint toxicity and their structural dissimilarity. To better understand these and other aforementioned results, we next turn to qualitative inspection of retweet content in our samples.

\begin{table}
\centering
\begin{tabular}{*{5}{c}}
\toprule
Clusters & $Granger_{T_{c1}\to T_{c2}}$ & $Granger_{T_{c2}\to T_{c1}}$ & $Granger_{T_{c1,c2}\to D_{c1,c2}}$ & $Granger_{D_{c1,c2}\to T_{c1,c2}}$ \\
\midrule
C1,C2 & 3.14e-01 (10) & 1.02e-01 (2) & 2.60e-04 (51) & 3.41e-02 (100)  \\
C1,C3 & 1.92e-03 (3) & 1.51e-03 (12) & 1.32e-02 (151) & 4.92e-04 (66) \\
C1,C4 & 1.04e-02 (155) & 6.07e-03 (4) & \textbf{2.17e-07} (43) & \textbf{9.86e-06} (27) \\
C2,C3 & 1.89e-02 (59) & 6.67e-03 (14) & 2.01e-03 (113) & 1.90e-02 (15) \\
C2,C4 & 4.29e-02 (68) & 1.17e-02 (2) & 1.56e-03 (40) & 8.00e-05 (1) \\
C3,C4 & 1.71e-03 (9) & 1.90e-02 (6) & \textbf{6.84e-06} (44) & \textbf{1.53e-08} (22) \\
\midrule
U1,U2 & 6.31e-04 (18) & 2.33e-03 (46) & 8.95e-06 (1) & 8.75e-05 (18) \\
U1,U3 & 1.84e-02 (1) & 4.50e-02 (49) & 1.58e-02 (1) & 3.14e-02 (26)  \\
U1,U4 & \textbf{1.02e-06} (1) & \textbf{9.61e-08} (1) & \textbf{1.85e-06} (1) & \textbf{1.22e-06} (1) \\
U1,U5 & 3.77e-02 (1) & 5.51e-03 (1) & 1.53e-02 (68) & 5.55e-02 (1) \\
U2,U3 & 1.30e-01 (83) & 3.29e-02 (1) & 1.47e-01 (1) & 1.01e-01 (1) \\
U2,U4 & 7.48e-03 (1) & 1.53e-01 (1) & 1.63e-03 (52) & 6.74e-03 (30) \\
U2,U5 & 3.35e-04 (16) & 3.34e-02 (28) & 1.11e-04 (24) & 8.66e-05 (16) \\
U3,U4 & 1.14e-01 (40) & 2.33e-03 (78) & 1.91e-02 (7) & 7.06e-02 (31) \\
U3,U5 & 8.51e-02 (83) & 2.25e-02 (50) & 7.22e-04 (8) & 1.45e-01 (6) \\
U4,U5 & 2.99e-03 (41) & 1.10e-03 (41) & 7.93e-03 (83) & 6.24e-02 (41) \\
\bottomrule
\end{tabular}
\caption{Test results for Granger-causality. First two columns report results for Granger-causality between toxicity prevalence in each cluster. Third and fourth columns report results for the Granger-causality between a cluster pair's structural dissimilarity and the likelihood of encountering toxicity in either cluster. Individual test p-values are bold when equal to or below the Bonferroni-corrected significance threshold (8e-6), with lags corresponding to the reported p-values indicated in parentheses.}
\label{granger_table}
\end{table}

\subsection{Interpretation}

Across these clusters, highly retweeted posts are the core metric for determining the influence of a poster. From observation, there are two distinct practices that result in this influence. The first is having a relatively small number of posts which gain a large number of retweets. This is common for politicians, celebrities, and influential journalists whose posts individually garner a great deal of attention. The second is posting a great number of times, with each post having a modest number of retweets but with a similar aggregate value to the celebrity influencers. This second group communicates more similarly to dialogue than the monologues of the celebrities. While our model does not distinguish between these two groups, it is an important characteristic of these clusters to keep in mind.

\subsubsection{Covid Clusters}

\begin{figure}[h!]
    \centering
    \includegraphics[width=0.85\textwidth]{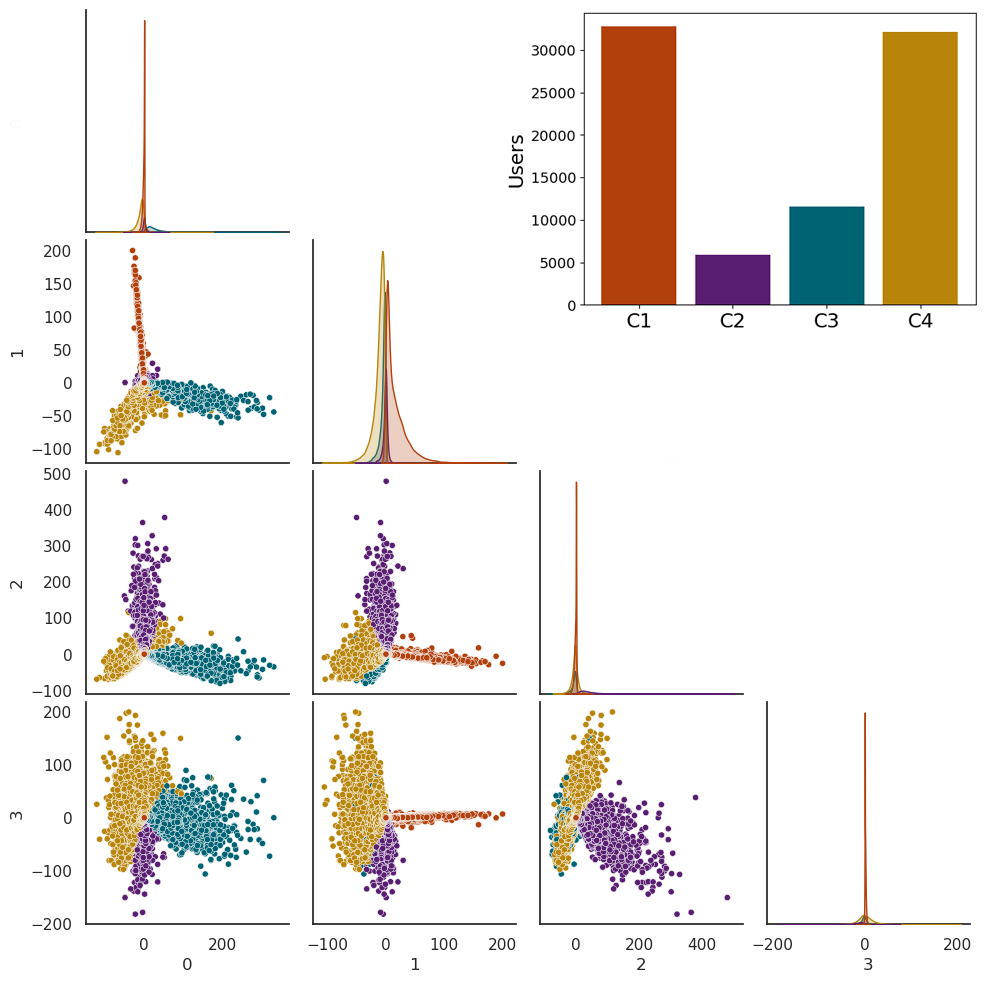}
    \caption{Left: Principal Component plots for the four retained components in the Covid sample. Color encodes users' cluster membership. Right: Bar plot showing cluster sizes. Colors in bar plot map to clusters in the PC plots.}
    \label{fig:cov_clust}
\end{figure}

As can be seen in Figure \ref{fig:cov_clust}, the first component in the Covid sample space separates out clusters according to the American and United Kingdom spheres of vaccination discourse, with Canadian vaccine mandate opposition and pro-mandate users situated between these two spheres. The second component, however, distinguishes clusters with pro- and anti-mandate positions, especially differentiating pro-mandate users in cluster C1 from the American anti-mandate cluster C4. The third component pulls out behavior specific to Canadian anti-mandate discourse especially in cluster C2. Lastly, the fourth component appears to correspond to differences between American and Canadian anti-mandate discourses. Although some components have clear geographic interpretations, they may simultaneously reflect important ideological and mandate-contextual differences that correlate with region.

\begin{figure}[h!]
    \centering
    \includegraphics[width=1\textwidth]{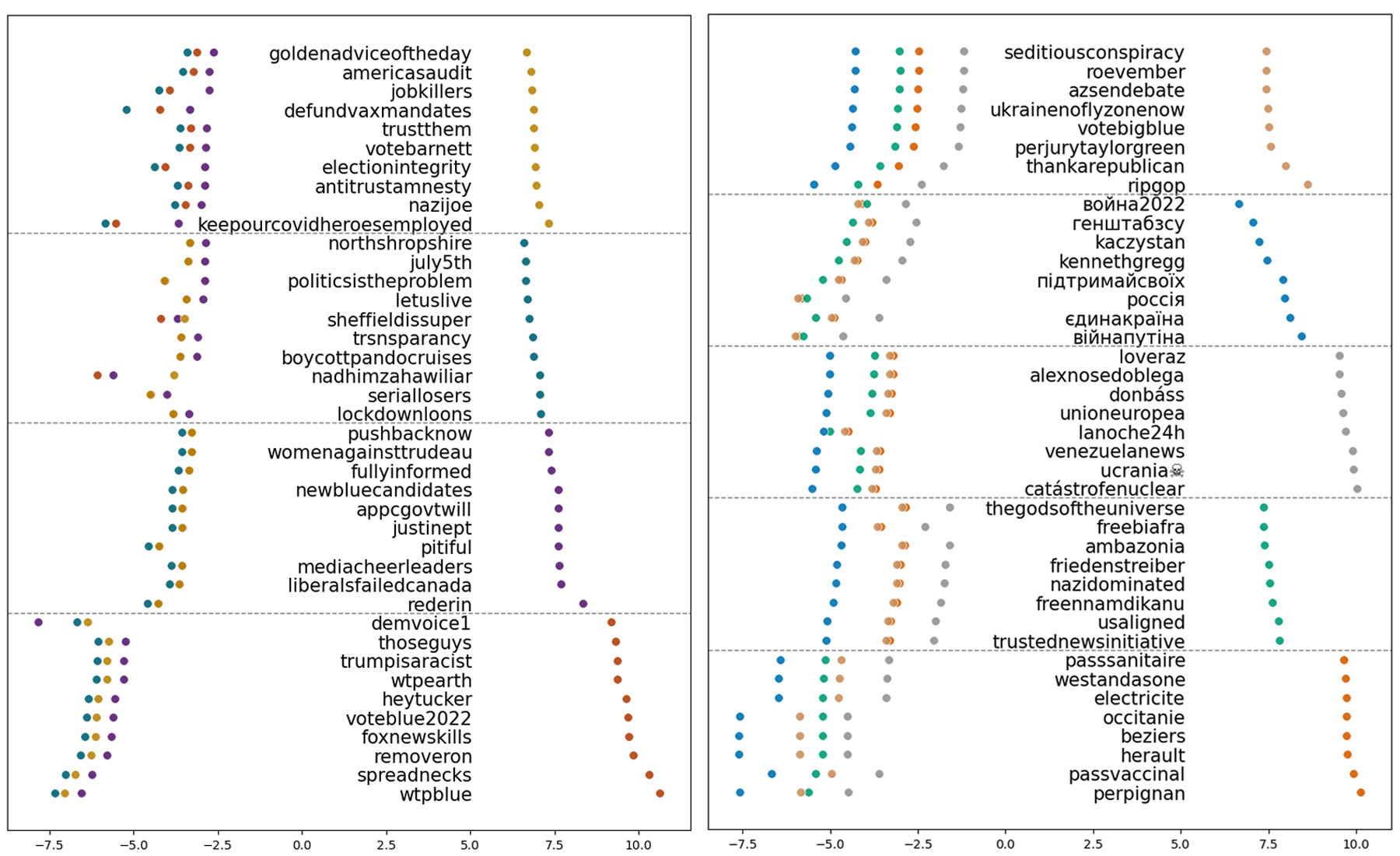}
    \caption{Hashtags with the top log-odds ratios per cluster (x-axis) for Covid (left) and Ukraine (right) samples. The log-odds ratio of a hashtag is computed in terms of the prevalence of that hashtag in a cluster compared to that hashtag's prevalence among users in all other clusters. Hashtag sets are ordered by cluster with the hashtags most idiosyncratic for C1 or U1 at the bottom and those for C4 or U5 at the top. Ratio markers are color encoded by cluster membership as specified in Figure \ref{fig:cov_clust} and Figure \ref{fig:ukr_clust}.}
    \label{fig:logodds}
\end{figure}

Cluster C1 similarly stands out in Figure \ref{fig:logodds}, with its more distinct hashtags listed at the bottom of the left-hand plot. The log-odds ratios for its most distinct hashtags are greater than for other clusters, highlighting its idiosyncrasy with respect to the other clusters. This is also observed in the substance of the hashtags, which contain a clear ideological bent, especially the hashtags ``trumpisaracist'' and ``foxnewskills.'' In contrast to C1, the rest of the clusters, while distinct by region, express similar ideological positions that share an opposition to various Covid public health policies. 

C1 and C4 stand as the most directly opposed in ideology. While both contain content referencing non-American Covid news and politics, the American context predominates. As seen in Table \ref{covid_examples}, C1 and C4 can display considerable toxicity towards each other. Despite this opposition, it would be a mistake to assume these clusters are consistently speaking coherently about each other. The presumed antagonists of C1 are often interpreted as being general anti-vaxxers, a more stigmatized label than opposition purely to the Covid-19 vaccines or mandates. C4 certainly contains general anti-vaccine sentiment, but (as is also true of C2 and C3) most of these messages are quite focused on the Covid-19 vaccine or policies like vaccine mandates and vaccine passports. Conversely, the presumed opponents of C4 are more often the elites: politicians, officials, and pharmaceutical companies than supporters of the vaccine campaigns. With both clusters focusing on opponents that are not representative of those speaking on the platform, non-polarized dialogue is difficult at best.

The relationships between C1 and both C2 and C3 are naturally somewhat reduced by the difference in country context. C1 and C2 share a fairly similar relationship to C1 and C4, since both the UK and Canada had administrations that implemented vaccine policies that members of C1 approve of and the other clusters disapprove of. Still, the American and Canadian experiences of Covid policies and lockdowns were more similar to each other than to those of other countries within the Anglosphere. In contrast, C3, representing mainly British anti-vaccine positions, displays an opposition to vaccine mandates and passports informed by the British experience of lockdowns. Moreover, C3 lacks the clear partisanship present in C1, C2, and C4. The Conservative governments of this period did not inspire a coherently partisan response in those opposed to their vaccine policies. 


The time series plots, as seen in Figure \ref{fig:cov_ts}, show the varying dissimilarity between the clusters in question over our time period as well as the toxicity. The black line present in each plot represents the start of the Ukraine war, after which point there is increasing instability in these clusters. The first and second halves of our dataset do display some distinctive differences. The prior descriptions of C2, C3, and C4 as being more interested in opposing vaccine passports than opposing the vaccines in principle is more true in the first half of our datasets, as more conventionally anti-vaccine content becomes more common later, as do references to other conspiratorial concepts like the Great Replacement or QAnon terminology. 

This is especially true for C4, which witnesses a spike in toxicity shortly after the start of the Ukraine war. This may in part be due to relatively moderate members shifting their focus to Ukraine. This increase in ideological content within C4 may explain why C3 and C2,  clusters that are more moderate in its opposition to mandates, starts to move away from C4 in the network structure, producing extreme values in the structural dissimilarity of C3,C4 and C2, C4. Although ostensibly aligned in their opposition to mandates, the movement of C3/C2 from C4 suggests that the latter becomes more ideological compared to C2 and C3. This observation raises the possibility of a distinct form of polarization dynamics characterized by gaps in how committed groups are to a common ideological framework.  

Within the first half of the time series there are several peaks in the structural dissimilarity and toxicity between C1 and the other clusters. The first peak occurs around April 1st, 2021 and coincides with a variety of government announcements following early efforts at mass vaccination. Vaccine passport policies are proposed, irritating C2, C3, and C4. But weaker than anticipated vaccination drives irritate members of C1. This amplifies the differences in discourse between these groups. At the same time, vaccine side effects became increasingly popular as a story, leading to increasing vitriol among those already suspicious of the vaccines.

The next spike in toxicity/structure in mid-July of 2021 is likely due to several causes. The growth of the Delta variant over the summer of 2021 led to some backtracking on easing restrictions in the US. Meanwhile, the UK mostly concluded its lockdown period. A popular subject within the posts from this period is debate over school policies for the 2021-2022 school year and whether schools should be open or not. In the US, several Republican-controlled states were publicly opposing various Covid measures like vaccine mandates.

There is another spike shortly before the Ukraine war disrupts this discourse in the last days of 2021. Dr. Robert Malone, a biochemist who promoted Ivermectin and hydroxychloroquine over the Covid vaccines was banned from Twitter on December 29th, 2021 and appeared on the Joe Rogan Experience podcast the next day. This event appears to have activated C1, C2, and C4 quite effectively, leading to a heightening of toxicity and structural drift while the story ran its course. 

\begin{table}
\centering
\begin{tabular}{p{0.1\linewidth} | p{0.4\linewidth} | p{0.4\linewidth}}
\toprule
Cluster & Example 1 & Example 2  \\
\midrule
C1 & How much of an idiot do you have to be to spend up to \$400 for a fake vaccine card, when you can get a REAL one for free & holy fucking shit, ``Donald Trump deserves endless praise for the vaccine I refuse to take'' is some industrial-strength batshit incoherence right there. \textbf{congratulations, morons} \\
C2 & The smoking gun…The Trudeau Liberals were dead wrong about their Covid-19 vaccine assumptions and policy. They knew they were wrong (since their own study told them in June). And, they kept moving ahead anyway. The Trudeau Liberals are despicable. & They’re alone in their trucks almost all the time. There was no crisis with sick truckers. YOU are the one forcing them off the road, disrupting the supply chain and endangering the food security of Canadians, absolute moron! \\
C3 & Blimey, I missed this last night, but I bet Tesco bosses are sweating right now. Who the hell thought it was a good idea to show a Vaccine Passport in their Xmas advert?! I’m not surprised \#BoycottTesco is trending. Idiotic. & This government is the purest evil. To think some of these scumbags were once my friends. I despise them all - and they will burn in hell for this \\
C4 & Let me get this straight... some Democrats want American citizens to have a Vaccine Passport to travel freely within the United States but not an ID to vote?!? Clowns!!! & People around the world are sick of the bullshit.', 'WE. WARNED. YOU!  Then you called us anti-science, anti-vax, horse paste eating conspiracy theorists who wanted to kill grandpa and grandpa. Epic fail for dumb, useful idiot, government sheep. \\
\bottomrule
\end{tabular}
\caption{Example posts for each cluster in the Covid sample. Example 1 posts are selected according to their popularity within a cluster and weighted by that post's toxicity. Example 2 posts give greater weight to toxic posts while ensuring that they are still relatively popular in that cluster.}
\label{covid_examples}
\end{table}

\subsubsection{Ukraine Clusters}

\begin{figure}[h!]
    \centering
    \includegraphics[width=0.85\textwidth]{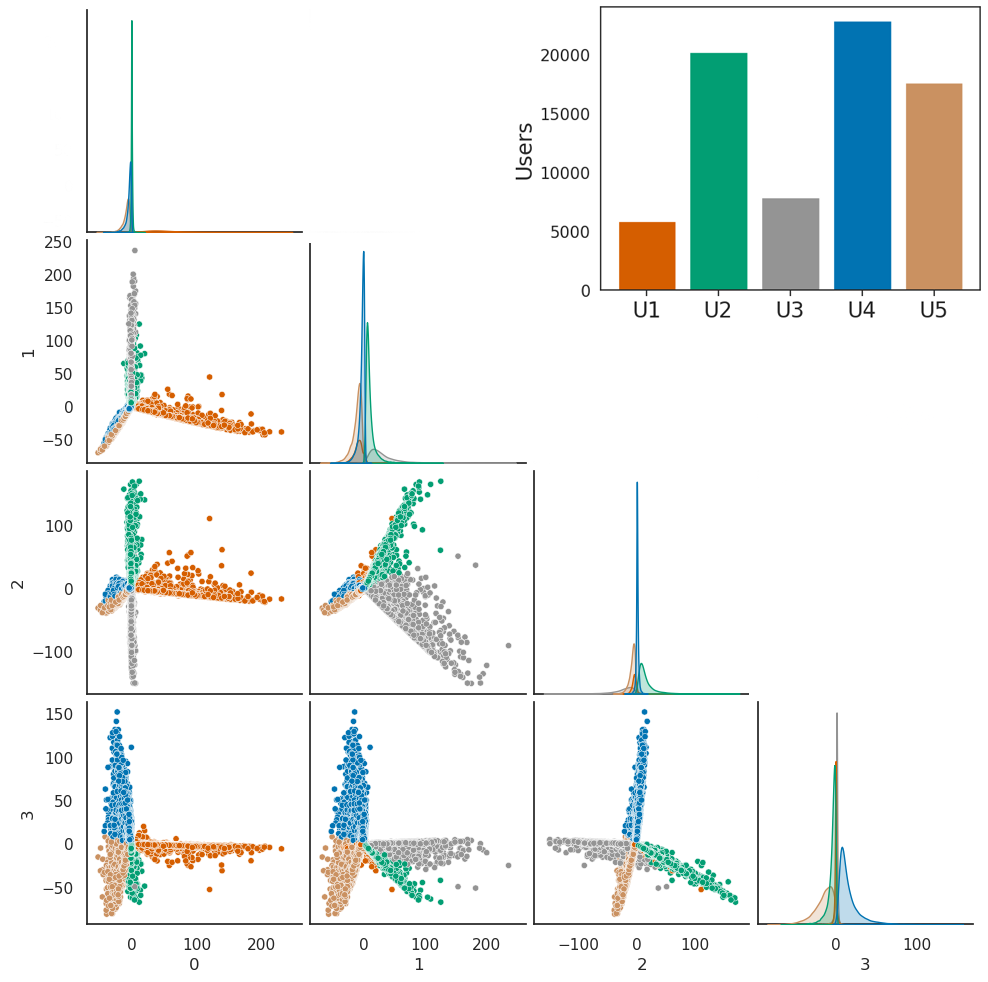}
    \caption{Left: Principal Component plots for the four retained components in the Ukraine sample. Color encodes users' cluster membership. Right: Bar plot showing cluster sizes. Colors in bar plot map to clusters in the PC plots.}
    \label{fig:ukr_clust}
\end{figure}

As seen in Figure \ref{fig:ukr_clust}, the first principal component pulls out U1, which is a French-speaking, pro-Russian cluster. The second component pulls the other pro-Russian clusters, U2 and U3, from the pro-Ukrainian clusters U4 and U5. The third component separates U2 from U3, containing English and Spanish speaking users, respectively. Lastly, the fourth component differentiates U4, an international pro-Ukrainian cluster, from American dominated pro-Russia and Pro-Ukraine clusters U2 and U5.

On the right side of Figure \ref{fig:logodds}, we can see the most distinct hashtags per Ukraine cluster. Due to being isolated linguistically, U1 and U3 (bottom and middle of the figure, respectively), the non-English language hashtags are easily separated from the other clusters. U4 sees hashtags in Ukrainian, identifying U4 as the cluster containing Ukrainian activists. Although predominantly English speaking, U4 also contain pro-Ukraine posts in German, French, and Spanish. U5, in contrast corresponds to an American perspective among pro-Ukrainian groups, as anti-Republican and/or 2022 midterm election hashtags are found here. U2's hashtags are more diverse, with African conflict hashtags alongside religious and anti-Ukrainian hashtags. Responsible for some of this diverse content, U2's activity contains activist Omali Yeshitela, who was charged and found guilty by the US DOJ of conspiracy to act as an agent of the Russian government (\cite{DOJ}).

In Table \ref{ukr_examples}, we see fairly complex examples of messages with high toxicity scores. These are often accompanied in the dataset by other posts of the format: ``<expletive> <name of politician(s)>.'' Toxicity is a complex concept for discussions of wars, as wars naturally invite words associated with hostility, confrontation, and violence. Still, rhetorical approaches to attacking opponents in a toxic fashion were not wholly distinct from those found in the Covid-19 dataset. 

The pro-Russian clusters (U1, U2, and U3) display general similarities. The French cluster U1 presents many of the pro-Russian argumentation found in the European context including spreading the Russian originating narrative that Ukriane has a bioweapons program. This position is often woven into criticism of the Macron government. U2 displays a variety of arguments aimed at a number of English speaking audiences. Some raise the cost of Ukrainian aid to a particular NATO country with that country as the assumed audience. Others defend Russia's invasion to a more global audience by denigrating Ukraine. Still others take a leftist, anti-Western position. Aside from fitting into Russia's narrative of the conflict, there is little rhetorical consistency even within the same month. Another element in U2 is the instruction to ``do your own research,'' while listing names or organization to search for that will likely lead to the author's intended conclusion. U3 contains Russian-aligned discourse in Spanish centered on Venezuelan discourse.

The pro-Ukrainian clusters (U4 and U5) are best differentiated by U4 focusing on the war from Ukraine's perspective while U5 focuses on American support for the war. A post discussing German aid in U4 was likely written with an international frame of reference as compared to U5 where aid is almost always framed in the context of partisan politics, especially during the US midterm elections. Many of the posts in U4 are news of some sort from a pro-Ukrainian perspective, whether about activity in the war, international responses to the war, or discussions about the media responses to the war. If there is a general argument being made, it is that Ukraine is the victim and NATO should help Ukraine. The U5 cluster is rather more inward-looking. Even at the start of the conflict, where news predominates, the inconsistent attitudes towards Ukrainian aid within the Republican party, the stance of Fox News, and the position of Donald Trump occupy a great deal of attention within this cluster. Eventually, the US midterm elections focus attention even further.

\begin{table}
\centering
\begin{tabular}{p{0.1\linewidth} | p{0.4\linewidth} | p{0.4\linewidth}}
\toprule
Cluster & Example 1 & Example 2  \\
\midrule
U1 & LA CRAPULE ZELENSKY 13 août Il exige que l'refuse tout visa d'entrée à tous les citoyens russes «au nom des valeurs» 31 août On apprend qu'il loue 50.000 € mois à un couple RUSSE la villa italienne somptueuse qu'il a achetée avec des fonds détournés (cf.Pandora Papers) & Zelensky a obtenu la nationalité anglaise. Ses parents viennent d\'obtenir la nationalité israélienne. Pendant ce temps ``so'' peuple souffre principalement des horreurs commises par son armée. Et bien sur, cest Poutine lenculé. Jen veux vraiment aux gens dêtre aussi cons...\\
U2 & Ukraine executed the false flag attack on Poland so poorly that there was no chance to blame Russia without losing all credibility. Now the west is calling it ``stray missiles'' and an ``accident'' adding that none of this would have happened if Russia didn’t invade Ukraine. Idiots. & It\'s not okay for grown adults to call the invasion of Ukraine ``unprovoked''; that\'s a fartbrained fairy tale for idiots and children. You have a whole internet full of information at your fingertips. Fucking use it. \\
U3 & Un tipo que se somete a una sesión de fotos así - en plena guerra en su país - sólo puede ser un demente o un payaso pagado para cumplir su papel. Zelenski es ambas cosas. &  Que jodan al gobierno alemán, que jodan a la Unión Europea y que jodan a todos los falsificadores de la historia y a todos los censores, porque fue la Bandera Roja de los Soviets, con la hoz y el martillo, la que ondeó orgullosa sobre el Reichstag el Día de la Victoria. \\
U4 & Russia has a hunger plan. Vladimir Putin is preparing to starve much of the developing world as the next stage in his war in Europe. &  Fuck your thoughts and prayers, Ukraine needs military aid. \\
U5 & NEW: Senator Roger Marshall (R-KS) just tried to attack President Biden on Ukraine by saying, ``Get them the damn weapons.'' Marshall just voted along with 30 other Republican senators against aid for Ukraine, including weapons. & holy fucking shit, if you're rooting for chaos in Ukraine because you think it makes Joe Biden look bad, kindly resign from Congress and slither the fuck back under your flat rock  \\
\bottomrule
\end{tabular}
\caption{Example posts for each cluster in the Ukraine sample. Example 1 posts are selected according to their popularity within a cluster and weighted by that post's toxicity. Example 2 posts give greater weight to toxic posts while ensuring that they are still relatively popular in that cluster.}
\label{ukr_examples}
\end{table}

The time series in Figure \ref{fig:ukr_ts} can give us indications of particular moments of dissension or toxicity between the clusters. A reasonably common spike in toxicity across our clusters occurs around the beginning of April 2022, which coincides with the release of photos and video of the Bucha massacre. Considering the violent and sexual language needed to merely describe the Bucha massacre, it is unsurprising that measured toxicity would increase, especially among the pro-Ukrainian clusters.

The next major spike comes between U1 and U2 on May 8th. This is likely due to VE Day celebrations in France and Russia, which may have resulted in a spike in structural dissimilarity. Moreover, while the posts in U1 highlight the history of French cooperation with the Soviets against the Nazis to suggest support for Russia in Ukraine, on the same day posts in U2 speculate about Ukrainian false flag attacks on the Victory Day celebrations the next day in Moscow.


A final major burst in structural dissimilarity and toxicity comes from the build-up and resolution of the September 6th Kharkiv counteroffensive, as well as concern about damage to the Zaporizhzhia nuclear power plant around that same time. Concern about nuclear fallout is found in the French cluster U1. While U1 places the blame for this crisis on Ukraine, U4 takes the opposite stance, placing responsibility on Russia. The geographic scope for concern about nuclear fallout may also have driven the structural divergence between the two pro-Ukrainian clusters U4 and U5.

\section{Discussion}

Our results speak to a degree of interactional polarization taking place in some of our cluster pairs with an important role played by users' affective signaling. For the most part, results fit with what we would expect. Both statistical and qualitative analyses of the American pro- and anti-mandate pairs point to a polarization process between these groups. The other statistically significant finding in the Covid sample is somewhat more unexpected. Despite nominally being anti-vaccine, the British and American clusters' structural dissimilarity and toxicity were found to be predictive of each other. On this point, we are limited to speculation, but an answer that fits in with the qualitative observations is that the British cluster was less partisan than the American anti-mandate cluster. Part of this may have been a reluctance by UK Conservatives to signal opposition to a government of the same party using emotional language. Another contributing factor may have been an overall greater propensity toward toxic language by American users as moderately suggested by Mann-Whitney tests results for cluster-wise toxicity distributions. The structural divergence of C2/C3 from C4 may potentially also indicate a form of internal polarization taking place. Future work may benefit from examining polarization processes occurring within a single ideological camp to evaluate if thinking in terms of ``commitment'' polarization is justified.

The situation in the Ukraine dataset is more complex for studying polarization due to greater geographic and linguistic cleavages. Our statistical analysis suggests a temporal dependency between the French pro-Russian cluster U1 and pro-Ukraine cluster U4. Although a candidate explanation for structural dissimilarity between these two clusters could be the differences in language, U4 also contains French language content, suggesting the two clusters' participation in a pan-European debate over Ukraine. This interpretation is supported by the opposing stances taken on the question of responsibility for the crisis over the Zaporizhzhia nuclear power plant. Another contributing factor may be Macron's role as one the leading pro-Ukraine European leaders, resulting in polarization among French users with implications for other pan-European partisan issues such as economic policy. 


\section{Limitations}

Although data from social media platforms provides new avenues for research on social phenomenon such as polarization, there are a number of challenges to conducting this research using traces of social media activity. Common limitations which the present analysis is also subject to include disproportionate focus on the American context. This comes about due to language skill limitations, as well as the central role American user bases occupy on platforms such as Twitter/X. 

Other challenges arise from a gap in established methodologies to measure the dynamics of polarization, or the ability to treat polarization as a multitude of sub-processes that may reinforce one another and together produce increasingly fragmented societies. In this study, we attempt to move in the direction of studying polarization as a set of dynamic processes, which in the context of social media, we approach by measuring preferences for (ideological) influencers as reflected in network structure and the propensity to amplify affect-laden posts.

In addition to needing further progress in this direction, many other challenges and limitations remain. Chief among these is that any study of a single or limited set of discussions on a single platform will be blind to how polarization dynamics are playing out on other platforms and other spheres of society, be it occupational, political, ethnic, or generational. Polarization dynamics may be occurring in parallel within multiple spheres and it is not always clear where the main driving force lies. When we observe dynamics on a single platform, it is also not clear if what we observe is truly reflective of a process occurring on that platform or if what we observe is simply a reflection of dynamics mostly occurring elsewhere. 

There is also a great deal of noise and other complexities captured by polarization proxy variables. Discussions on social media often straddle multiple topics. Social networks inferred from social media posts are often incomplete and correspond to different types of user relationships coexisting in a single network, resulting in interpretability challenges. Using toxicity to measure affective signaling is challenging due to the uncertainty surrounding how well a chosen model handles toxicity inference for different communication types. Bias can be introduced when dealing with multiple languages or dialects. In addition, British and American standards for swearing can lead to differing evaluations of post toxicity. Perhaps most challenging, coded language such as ``Let's Go Brandon'' may be important to polarization dynamics, but difficult to identify and measure through quantitative approaches.

\paragraph{Acknowledgments}

We are grateful for helpful suggestions made by Alessandro Flammini and Filipi Nascimento Silva, as well to numerous others who participated in the original data collection and helped with research infrastructure.

\paragraph{Funding Statement}
This study was supported in part by a grant from the Knight Foundation to the Indiana University Observatory on Social Media (OSoMe). This work was also enabled by Jetstream2 at Indiana University from the Advanced Cyberinfrastructure Coordination Ecosystem: Services and Support (ACCESS) program, which is supported by National Science Foundation grants \#2138259, \#2138286, \#2138307, \#2137603, and \#2138296.

\paragraph{Competing Interests}
None

\paragraph{Data Availability Statement}
Data will be made available to other researchers upon request.

\paragraph{Ethical Standards}
The research meets all ethical guidelines, including adherence to the legal requirements of the study country.

\paragraph{Author Contributions}
Conceptualization: D.A; J.P. Methodology: D.A; J.P; B.P.H. Data curation: D.A. Data visualisation: D.A. Writing original draft: D.A.; J.P.; B.P.H. All authors approved the final submitted draft.



\printbibliography



\end{document}